\documentclass[modern]{aastex631}     

\newcommand{\etal}{et al.}

\usepackage{graphicx}
\received{May 19, 2022}
\revised{June 6, 2022}
\accepted{ June 9, 2022}
\published{August 4, 2022}
\submitjournal{ApJ}

\begin{document} 
\title{The Great Dimming of Betelgeuse: a Surface Mass Ejection (SME) \\ and its Consequences}
\shorttitle{Betelgeuse Great Dimming}

\author[0000-0002-8985-8489]   {Andrea K. Dupree}
\affiliation{Center for Astrophysics $|$ Harvard \& Smithsonian, 60 Garden Street, MS-15, Cambridge, MA 02138, USA}
\author[0000-0002-6192-6494]{Klaus G. Strassmeier}
\affiliation{Leibniz-Institut f\"ur Astrophysik Potsdam (AIP), An der Sternwarte 16, D14482 Potsdam, Germany}
\author{Thomas Calderwood}
\affiliation{American Association of Variable Star Observers, 49 Bay State Road, Cambridge, MA 02138}
\author {Thomas Granzer}
\affiliation{Leibniz-Institut f\"ur Astrophysik Potsdam (AIP), An der Sternwarte 16, D14482 Potsdam, Germany}
\author{Michael Weber}
\affiliation{Leibniz-Institut f\"ur Astrophysik Potsdam (AIP), An der Sternwarte 16, D14482 Potsdam, Germany}
\author{Kateryna Kravchenko}
\affiliation{Max-Planck Institut for Extraterrestrial Physics, Giessenbachstrasse 1, D-85748 Garching, Germany}
\author [0000-0002-3728-8082]{Lynn D. Matthews}
\affiliation{Massachusetts Institute of Technology, Haystack Observatory, 99 Millstone Road, Westford, MA 01886 USA}
\author[0000-0002-7540-999X]{Miguel Montarg\`es}
\affiliation{LESIA, Observatoire de Paris, Universit\'e PSL, CNRS, Sorbonne Universit\'e, Universit\'e Paris Cit\'e, 5 place Jules
Janssen, 92195 Meudon, France}
\author{James Tappin} 
\affiliation{RAL Space, STFC Rutherford Appleton Laboratory, Harwell Campus, OX11 0QX Didcot, UK}
\author{William T. Thompson}
\affiliation{ADNET Systems Inc.,  NASA Goddard Spaceflight Center, Code 671, Greenbelt, MD 20771}

\begin{abstract}

The bright supergiant, Betelgeuse (Alpha Orionis, HD 39801), underwent 
a historic  optical dimming during 2020 January 27 $-$ February 13. Many imaging and spectroscopic observations across the 
electromagnetic spectrum were obtained 
prior to, during, and subsequent to this dimming event. 
These observations  of Betelgeuse
reveal that a substantial surface   
mass ejection (SME) occurred  and moved out through the
extended atmosphere of the supergiant.    A photospheric shock occurred in 2019 January - March, progressed through the 
extended atmosphere of the star during the following 11 months and led to  dust production in the atmosphere.   Resulting from the 
substantial mass outflow,  the stellar photosphere 
was left with lower temperatures and the chromosphere with a lower density.      The mass ejected  could represent a 
significant  fraction of the total annual mass loss rate from the star suggesting that episodic mass loss events 
can contribute an amount comparable to that of the stellar wind.   
Following the SME,  Betelgeuse was left  with a  cooler average photosphere,  an unusual short photometric oscillation, reduced
velocity excursions,   and the disappearance of the $\sim$400-day pulsation in 
the  optical and radial velocity  for more than 
two years following the Great Dimming.
\end{abstract}

\section{Introduction}

Betelgeuse has been observed visually or photometrically for more than a century
revealing its semi-regular variability.  This red supergiant star is nearby and large in apparent size providing a 
unique opportunity for intensive  and 
spatially resolved study of its atmosphere and surroundings - in addition to its potential as a supernova. 
Two principal periods of  light variation occur
in the star: $\sim$400 days believed due to pulsation in the fundamental mode (Joyce et al. 2020) and 
a longer secondary period of $\sim$ 5.6 years 
(Kiss et al 2006; Chatys et al. 2019; Stothers 2010), typical of red supergiants but currently of uncertain origin (Joyce et al. 2020).  During early December 
2019, the visual magnitude of Betelgeuse  became fainter than V $\sim$1 (Guinan et al. 2019).  The stellar magnitude reached
a historic  minimum V-magnitude of $\sim$1.65 between 2020 January 27 $-$ February 13, and subsequently recovered (AAVSO 2022).   The 
star's appearance changed dramatically
in  late 2019 December and the early months of 2020, becoming substantially darker over the southern hemisphere in optical light 
(Montarg\`es \etal\  2021).   

The following sections  describe the various observations of Betelgeuse, including new temperature diagnostics and velocity measurements.
  We conjecture that a surface mass ejection (SME) occurred and present  recently observed 
atmospheric consequences for the star.

\medskip

\section{Betelgeuse: Fundamental Parameters}

Decades of observations of Betelgeuse have resulted in hundreds of measurements; a summary of  recently derived  fundamental parameters is given 
in Table 1.  Current technologies, especially interferometric techniques determine the apparent photospheric diameter as 42 mas which is  within the errors
of the pioneering measurement by Michelson and Pease (1921) of  47$\pm$4.7 mas.  Determination of the  stellar distance  from parallax measures remains 
challenging.  Because the star is large in apparent size, and the surface contains variable bright convective cells, the measurement of a parallax based 
on identification of the center without 
contemporaneous knowledge of the brightness distribution can produce  discrepant results (van Leeuewen 2007; Chiavassa \etal\ 2022).  For the estimates
following in later sections,  we adopt an apparent diameter of 42 mas (Ohnaka \etal\ 2011), a  
value of 222 pc for the distance (Harper \etal\ 2017), and a radius of 1000R$_\sun$.  The stellar effective temperature can be measured using a variety of
diagnostics.  We select a `nominal' value of 3650$\pm$50K derived through a fitting of MARCS models to spectrophotometry of the star (Levesque \etal\ 2005)
to represent the average normal effective temperature. 

Subsequent sections report many determinations of the effective temperature of Betelgeuse as the Great Dimming occurred.  Analysis of the TiO molecule where signatures  appear in the optical and the infrared
regions has been employed in many ways to infer T$_{eff}$: high-resolution spectra, moderate resolution spectra, photometric measures.  These spectra
are generally modeled by assuming a grid of temperatures in the models to obtain the best agreement with observations.  A 
detailed discussion of different procedures is given by Kravchenko \etal. (2021) where these authors conclude that the bands at 6187\AA, 7085\AA. 
and 7125\AA\  are to be preferred because they
exhibit the most sensitivity to temperature.  Another technique involves the ratio of  line strengths between   two photospheric lines
of different excitation potential, V I and Fe I.  While this was calibrated for giant stars (Gray \&  Brown  2001), it can indicate a relative change in 
temperature of supergiants (Gray 2008), and is discussed in the following sections.  Millimeter and centimeter observations are employed as well to determine the electron temperature in the outer
atmosphere.    The radio continuum in these  bands results 
predominantly  from free-free  emission.  Because it is thermal and optically thick, a flux  measurement allows evaluation of the mean electron temperature 
(Reid \& Menten 1997).   Measures at 7-mm in 1996  suggested that the flux  arises from cooler components in an inhomogeneous extended 
atmosphere (Lim \etal\  1998),  and this result is in harmony with measurements in 2019-2020.   

Periodic variations in V magnitude and radial velocity are present.  A short period of $\sim$400 days is well-documented. This 
period is believed to represent the fundamental pulsation mode of the supergiant (Joyce \etal\ 2020). The phase of the radial velocity 
variation  lags the phase of the V magnitude variation by $\sim$ 35 days (Granzer \etal\ 2021).   The source of the much longer ($\sim$2100 days)
secondary period is not established.  It may represent the characteristic turnover time of giant surface convective cells (Stothers 2010).

\floattable

\begin{deluxetable}{lclc}


\tablecaption{Fundamental Parameters of Betelgeuse\tablenotemark{a} \\}
\tablehead{
\colhead{Property}& \colhead{Value}&\colhead{Comment}& \colhead{Reference} \\
}
\startdata
Temperature & 3650$\pm$50 K & Spectrophotometry + MARCS Models & 1  \\
Spectral Type &  M1-M2Ia-Iab  & Photographic spectra  & 3  \\
Radial Velocity  & $+$21.91$\pm$0.51 km s$^{-1}$  & Integrated photospheric spectrum   &  2  \\
Diameter  & 42.05$\pm$0.05 mas   & 2.28$-$2.31 $\mu$m, uniform disk    & 13  \\
Diameter&43.15$\pm$0.50 mas & 1.65$\mu$m (H-band) limb-darkened disk  & 14\\
Diameter& 125$\pm$5 mas & 2500\AA\ continuum & 15\\
Distance  &  153  $+$22, $-$17 pc & Hipparcos revised  & 4  \\
Distance &  168.1 $+$27.5, $-$14.9 pc & Seismic analysis & 5 \\
Distance &  222  $+$48, $-$34  pc & Hipparcos rev. + radio positions & 6 \\
Radius   &   764  $+$ 116, $-$62 R$_\sun$  & Seismic analysis      &   5 \\
Radius   & 996$+$215, $-$153 R$_\sun$ & Using Hipparcos rev. distance + radio & 6\\
Optical Variability & $\sim$ 0.3$-$1.2   &  Visual magnitude: 1917$-$2005  & 10 \\
Periodicity (fundamental) & 388$\pm$30 days& Using Visual magnitude: 1917$-$2005 &10\\
Periodicity (fundamental)& 420 days & Using B mag., UV cont., Mg II:1984$-$1986 & 11\\
Periodicity (fundamental)& 385$\pm$20 days&Using radial velocity: 2008$-$2022 & 11\\
Secondary Period & 2050$\pm$460 days & Using Visual magnitude: 1917$-$2005 &10\\
Secondary Period & 2229.8$\pm$6 days & Using radial velocity: 2008$-$2022 & 11\\
Position angle North Pole &  48.0$\pm$3.5$\arcdeg$   & ALMA measurement   & 8 \\
Position angle North Pole &  55$\arcdeg$ &Ultraviolet measurement  & 9\\
\enddata
\tablenotetext{a}{Extensive tables of measured parameters  dating back to over a century are
contained in  Dolan et al. 2016.} 
\tablerefs{(1) Levesque et al. 2005; (2) Famaey et al. 2005; (3) Keenan \& McNeil 1989;
(4) Van Leeuwen 2007; (5) Joyce et al. 2020; (6) Harper et al. 2017; (7) AAVSO 2022;
(8) Kervella et al. 2018; (9) Uitenbroek et al. 1998; (10) Kiss et al. 2006; (11) Granzer et al. 2021;
(12) Dupree et al. 1987; (13)Ohnaka et al. 2011; (14) Montarg\`es et al. 2016;
(15) Gilliland \& Dupree 1996
}

\end{deluxetable}

\section{Betelgeuse in 2019-2020: the Origin of the Great Dimming}
\subsection{The First Phase: 2019 January - November}

The sequence of multifrequency observations 
is detailed in Table 2, including estimates of formation level of the radiation in the atmosphere of Betelgeuse.
On  2019 January 1, images of the star in the optical continuum from VLT/SPHERE  revealed
a roughly symmetric stellar image (Montarg\`es et al. 2021).
Tomographic study, using high-resolution HERMES optical spectra of the unresolved star (Kravchenko et al. 2021; Raskin et al. 2011) 
allows a probe of the dynamics of several levels in the photosphere of the supergiant.   And during  2019 January,
the photospheric layers exhibited the velocity signatures of a strong shock in the line of sight, and an expansion of
the photospheric layers through 2019 March.

The photospheric velocity was measured continuously during this time  by the automated
STELLA telescopes (Granzer et al. 2021) and demonstrated that the  photosphere, began  to expand  in 2019 January  and 
maintained a constant maximum outflowing  value of $-$8  km s$^{-1}$  for most of the year until 2019 mid-November, when the outward velocity
began to decrease.

Optical spectra were obtained with the Tillinghast Reflector Echelle Spectrograph (TRES) mounted on the 1.5m-telescope
  at Fred Lawrence Whipple Observatory on Mt. Hopkins, AZ.  The Ca II K-line in the lower chromosphere exhibited  asymmetric emission wings signaling
outflowing plasma late in 2019 January and early March (Fig. 1). The extended chromosphere appeared undisturbed  
through  early 2019, as spatially resolved UV spectra appeared similar to one  
other,  and a `normal' (unenhanced)  distribution of the chromospheric Mg II flux and ultraviolet continuum  across the image  from 2019 January 
to 2019 March  5 (Dupree et al. 2020b). 

From 2019 April to 2019 August, signatures of atmospheric cooling were detected.  An increase in the 6$\mu$m band indicative of 
increased  H$_2$O formation (Taniguchi et al. 2021, 2022), 
and both mm and cm VLA images suggest exceptionally low temperatures, $\sim$ 2400K (Matthews \& Dupree 2022).

\floattable
\begin{deluxetable}{llclc}
\tabletypesize{\footnotesize}
\tablecaption{Spectroscopic and Imaging Observations of Betelgeuse 2019-2020 \\}
\tablehead{
\colhead{Date}& \colhead{Spectral Region}&\colhead{Formation}& \colhead{Observation}  & \colhead{Reference} \\
  &  &R$_{star}$ & & 
}
\startdata
2019 Jan  1 & Optical Image& 1.0& VLT/SPHERE-ZIMPOL Symmetric appearance  & 1\\
2019 Jan 1 & Optical Spectra& 1.0& STELLA: Radial velocity outflow begins & 16\\
2019 Jan-Apr  & Optical Spectra & 1.0 & Photospheric shock develops   & 2 \\
2019 Jan 25 + Mar 5 & UV Spectra\tablenotemark{a}& 1.1 - 3& Unenhanced Mg II UV fluxes  &3\\
2019 Jan 31 + Mar 5 & Ca K  Spectra& $\sim$1&Outflow - low chromosphere &15\\
2019 Apr 1-15 &6$\mu$m  Himawari&$\sim$1.1&H$_2$O increase & 14\\
2019 Aug 2   &  mm-VLA  Image&  $\sim$2.1 & low T (2267K) & 4\\
2019 Aug 2 & cm - VLA Image& $\sim$ 2.8 & low T (2583K) &4\\
2019 Sep 18$-$ Nov 28 & UV spectra\tablenotemark{a} & $\sim$1.1-3 &Enhanced chromosphere (Mg II) & 3 \\
  &  &  &Dense outflow/Southern hemisphere & \\
2019 Nov 4 - Dec 8 &Optical Spectra & 1.0& TiO temperature $\sim$3570K  &2\\
2019 Dec 7&TiO, near IR& 1.0&photometry: low T (3580K)&11\\
2019 Dec 27  & Optical Image & 1.0 & VLT/SPHERE-ZIMPOL: faint southern hemisphere & 1 \\
2020 Jan 17ff & Optical spectra & 1.0   &V I/Fe I implies T extremely low&15\\
2020 Jan 23  & Submm 450, 850 $\mu$m & $\sim$1  &low T (3450K)& 6\\
2020 Jan 28  &  Optical Image&  1.0 & VLT/SPHERE-ZIMPOL: faint southern hemisphere & 1\\
  &  &  &Dust and/or cool spot & \\
2020 Jan 31 & near IR Spectra & 1.0 & Te lower  3476K& 9\\
2020 Minimum & TiO: optical & 1.0 &Minimum value 3550 K & 2 \\
2020 Feb 3-Apr 1 & UV spectrum\tablenotemark{a} & 1.1-3& Chromosphere, Mg II  unenhanced &3\\
2020 Feb 4 & 10$\mu$ Himawari& $\sim$1.2 &Increased O-rich circumstellar dust&14\\
2020 Feb 8 + 19 & N band/MATISSE & 1.4 & Localized dust and/or cool spot & 12\\
2020 Feb 12 & Optical Spectra & 1.0 & Increased turbulence&10\\
2020 Feb 15  & TiO; near-IR filter & 1.0   & T=3520$\pm$25K  & 5 \\
2020 Feb 15  & TiO; optical spectra & 1.0  & T=3600$\pm$25K  & 7 \\
2020 Feb 17 & X-ray& ~1-2& non-detection& 13\\
2020 Feb. 23   &400-740 nm & 1 & aperture polarimetry; polarization from & 8\\
         & & & photosphere and/or obscuration by grains& \\
2020 Feb 20$ - $Apr 2020 &  UV spectra\tablenotemark{a} &1.1-3 &Lower density C II chromosphere/Southern hemisphere  &3\\
   & & &Southern hemisphere& \\
2020 Mar 20 & Optical Image & 1.0 & VLT/SPHERE-ZIMPOL   & 8 \\
\enddata
\tablenotetext{a}{Spatially resolved Ultraviolet spectra (7-8 pointings across surface) .}
\tablerefs{(1) Montarg\`es  et al. 2021; (2) Kravchenko et al. 2021;
(3) Dupree et al. 2020b; (4) Matthews \& Dupree  2022 ;
(5) Harper  et al. 2020; (6) Dharmawardena et al. 2020;
(7) Levesque \& Massey 2020; 
(8) Cotton et al. 2020; (9)Alexeeva et al. 2021;
(10) Zacs \& Pukitis 2021; (11) Guinan et al. 2019;
(12) Cannon et al. 2022; (13) Kashyap et al. 2020;
(14) Taniguchi et al. 2021; Taniguchi et al. 2022; (15) this paper; (16) Granzer et al. 2021}
\end{deluxetable}

\begin{figure}
\begin{center}
\includegraphics[angle=90, scale=0.6]{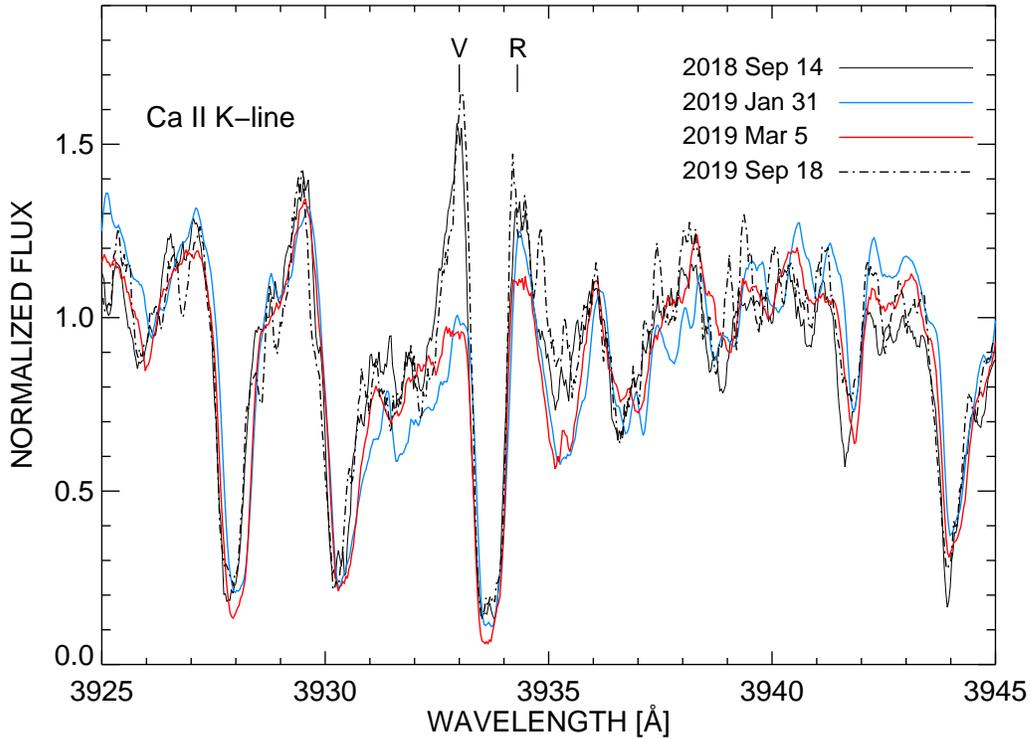}
\end{center}
\caption{The chromospheric Ca II  K-line, 3933\AA, from TRES echelle spectra,  displays asymmetric emission 
wings  [{\it marked as a V(violet) and R(red) feature}]. The appearance of V$\leq$R, signals wind absorption from a differentially 
expanding atmosphere in 2019 January and March.  The profile  lost this asymmetry in 2019 September, 
mimicing the earlier profile on 2018 September  14. }
\end{figure}

\begin{figure}

\includegraphics[angle=0,scale=0.55]{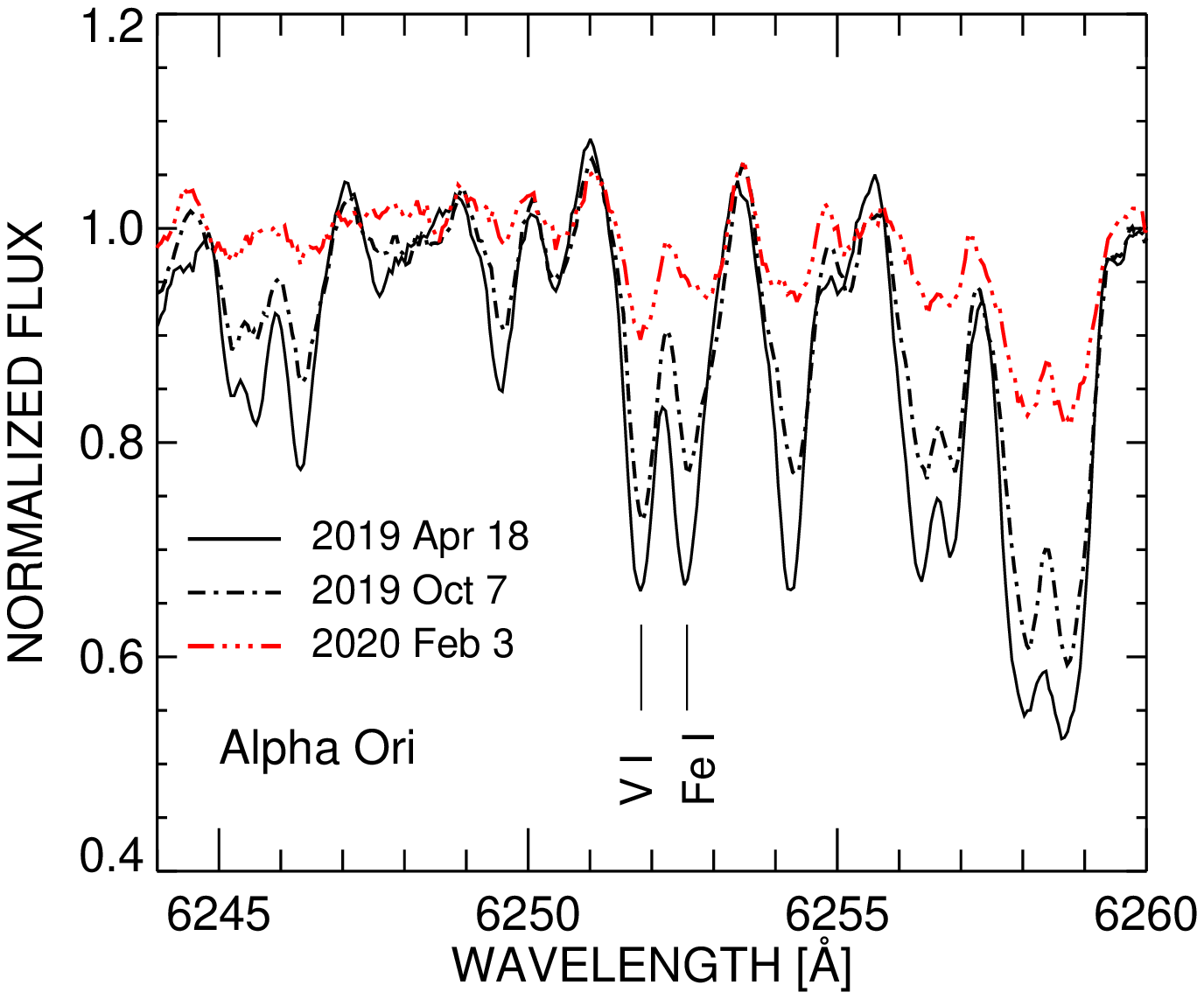}
\vspace*{-2.7 in}

\hspace*{+3.3 in}
\includegraphics[scale=0.45]{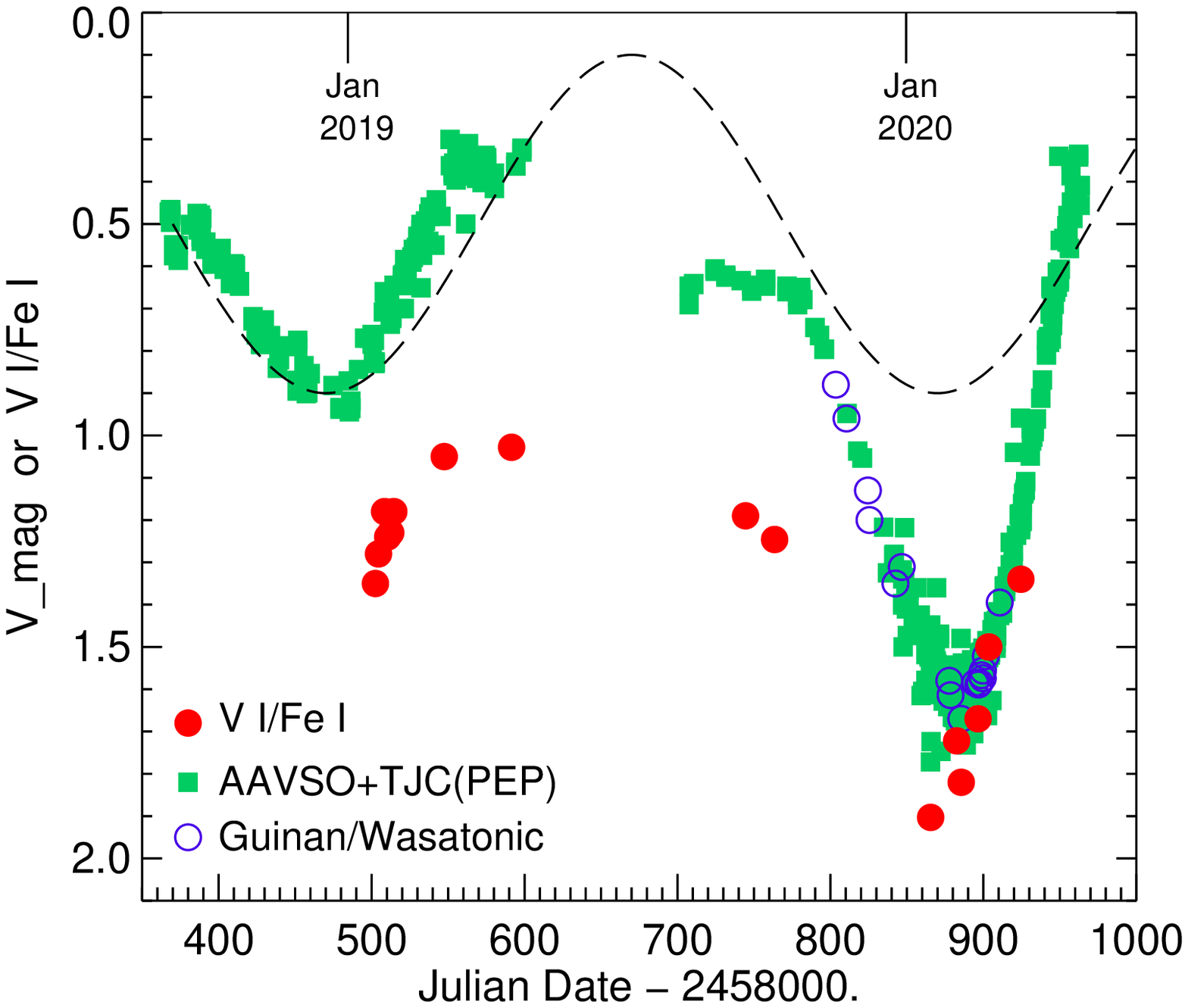}
\vspace*{+0.2in}
\caption{{\it Left panel:} Three TRES echelle spectra of the temperature sensitive  V I (6251.83\AA) and Fe I (6252.57\AA) absorption
lines from 2019 April 18 to 2020 February 3.  The Fe I line becomes weaker relative to the V I transition during this time span. Note
the broadening of the Fe I line on 2020 February 3 which appears to be blended at low T (see also Fig. 3).
{\it Right panel:} The V I/Fe I depth ratios clearly follow the behavior of the V-magnitude indicating a decrease of
the photospheric temperature.  The broken line marks a  $\sim$400 day period.}

\end{figure}

\smallskip
The shock in the photosphere moved out through the extended  atmosphere of the star in the  months following its 
emergence from the surface in 2019 March-April.   Here the term photosphere refers to  the surface which corresponds to 1 R$_\star$.  Assuming a
velocity of 5$-$10 km s$^{-1}$, the ejected material would reach a distance of 0.1$-$ 0.2 R$_{star}$  above the photosphere in 6 months (assuming R$_{star}$= 1000R$_{sun}$, 
Ohnaka  et al. 2011; Harper \etal\ 2017).   This is comparable to  the level where the contribution
function for ultraviolet Mg II emission becomes large in semi-empirical models  (Lobel \& Dupree 2000); moreover  Mg II
continues  to appear further out  in the atmosphere - up to 6 times the optical diameter (Uitenbroek et al. 1998; Dupree et al.  2020b). Assuming the
shock emerged from the photosphere in 2019 March-April,  a transit 
time of $\sim$ 6 months is in harmony with   the quiescent Mg II chromospheric  emission observed in 2019 March,
followed by  the enhanced Mg II emission which was spatially resolved  over the southern hemisphere  in 2019 mid-September  through 2019 November.   
Chromospheric emission lines of C II  demonstrated the 
increased chromospheric density during 2019 September $-$ November,
where and when  the Mg II emission was enhanced (Dupree et al. 2020b).  Such behavior signals a continuous outflow event of higher density material.

\subsection{The Historic Anomalous Phase: 2019 December - 2020 March}

In early 2019 December, photometry in the near IR suggested slightly lower temperatures ($\sim$3580K) were present in the photosphere than the
average temperature, 3650K (Levesque et al. 2005) or the value of  $\sim$3660K
found near a maximum brightness of V$\sim$ 0.2$-$0.3 mag (Guinan et al. 2019).
On  2019 December 27,  imaging with SPHERE-ZIMPOL  in the optical continuum near H-$\alpha$   revealed a dark southern hemisphere had appeared
on the star, creating a dramatic contrast with the previous image obtained on 2019   January 1  (Montarg\`es et al. 2021).  

By 2020 January 23, submillimeter spectra (Dharmawardena et al. 2020 ) indicated a temperature of $\sim$3450K in the photosphere,  
and IR spectra (Alexeeva et al. 2021) obtained 8 days later  confirmed a low  
effective temperature of the star  - this value being $\sim$3476K .   

Another  way to determine the change in the temperature of the photosphere relies on the  ratio
of two atomic lines, V I ($\lambda$6251.83) and Fe I ($\lambda$6252.5). These features indicate the photospheric
temperature in giant stars (Gray and Brown 2001), and were used to track the temperature in Betelgeuse itself (Gray 2000; Gray 2008; Weber et al. 2009).  Because the Boltzmann 
excitation factors differ, (0.29~eV for V I and 2.40~eV for
Fe I), they can serve as a temperature index.  The ratio, V I/Fe I becomes larger with decreasing photospheric temperature.  The line-depth ratio roughly followed the  
V-magnitude variations of Betelgeuse for about 12 years beginning in 
1995 (Gray 2008).  During that time the largest value of the ratio amounted to  $\sim$1.35 and corresponded to a temperature 
decrease of more than $\sim$ 100K.  
TRES spectra (Fig. 2) obtained at the Fred Lawrence Whipple Observatory in 2019-2020  allowed measurement of  this line depth ratio, and during the Great 
Dimming, the variation of this ratio corresponds well to the
V-magnitude of the star, and it  became very large (Fig. 2), reaching a value of 
$\sim$1.9.  This  value   suggests a larger temperature change than $\sim$250K.  This value  is a lower limit to T since the lines are not weak.  Moreover, it
appears that an adjacent line contributes to the Fe I feature at lower temperatures., suggesting that the ratio could become even larger.

The overall weakening of  absorption lines from 2019 April to  2020 February 3 as shown in Fig. 2, accompanied by the 
implied decrease in photospheric temperature
is supported by spectra of other supergiant stars (Fig. 3) where the T$_{eff}$ is known (Levesque et al. 2005).  For comparison, TRES echelle spectra obtained
of HD 156014 (T$_{eff}$= 3450K) and HD 175588 (T$_{eff}$= 3550K) reveal  a systematic weakening of the predominantly 
neutral absorption lines in this spectral region  with decreasing T$_{eff}$.

\begin{figure}
\begin{center}
\includegraphics[angle=90,scale =0.5]{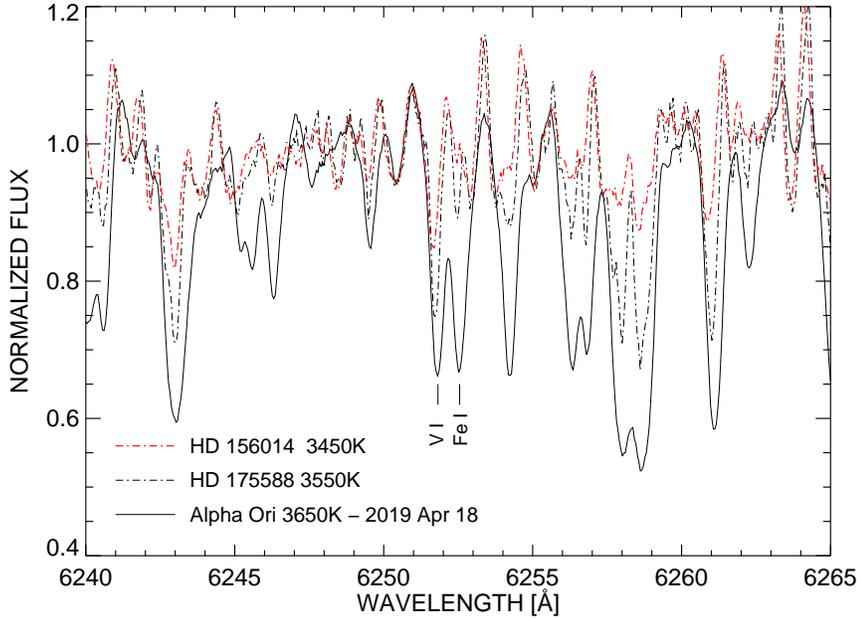}
\end{center}
\caption{TRES spectra of Betelgeuse in 2019 April, a cooler supergiant (HD 156014), and a bright giant star (HD 175588).  Note the systematic weakness
of the absorption features as the effective temperature  of the star decreases.  At lower temperatures, the Fe I line at 6252.55\AA\ becomes
blended with an unidentified absorption line.  At the lowest temperature (3450K), the unidentified line dominates the Fe I line,
making the ratio measurement  (V ~I/Fe~I) in Betelgeuse a lower limit when the star is faint.}

\end{figure}

If a large cool volume in the photosphere resulted from the surface mass ejection, both the size of the cool area and the temperature
differential from the rest of the atmosphere would contribute to a decrease in V-magnitude.  An estimate can be made of the
magnitude decrease at the maximum of the V-band from a temperature of 3650K  (Levesque et al. 2005), by assuming both 
lowered temperatures and various surface area coverage.   This 
shows that a  0.4 magnitude decrease occurs when 75\% of the star is covered by a patch 250K cooler than the remaining surface. 
A 0.5 magnitude decrease requires 90\% of the star to be cooler by 250K. Even if the cool  spot temperature drops
by 400K, 60\% of the star must be covered by a cool spot to achieve a 0.5 V-magnitude decrease.  And the photospheric temperature
measures cited in Table 2, do not approach 3250K.  During the Dimming, a value of 3450 K (Alexeeva et al. 2021) is reported 
 - a decrease of $\sim$200 K.  Direct imaging (Montarg\`es et al. 2021)  demonstrates  that at most 50\% of the star appears 
dimmer,  perhaps only 25\% as the dark area 
appeared in the southwest quadrant, suggesting that dust contributed to the dimming as others have suggested (Montarg\`es et al 2021;
Cannon et al. 2022; Levesque \& Massey 2020; Cotton et al. 2020; Taniguchi et al. 2022).

Now, from 2020 January 27 to February 13,  the V-magnitude reached a historical  
low level. However, on 2020 February 3 (and until 2020 April 1), the Mg II chromospheric emission and the ultraviolet continuum  had returned  
to levels commensurate with 2019 March which were not enhanced (Dupree et al. 2020b). Optical spectra obtained on 2020 February 12 
indicated increased turbulence in the photosphere (Zacs \& Pukitis 2021).   On 2020 February 15, somewhat different 
values of T$_{eff}$  were obtained with different diagnostics 
from the unresolved star.  Optical TiO transitions 
suggested the temperature to be 3600$\pm$25K (Levesque \& Massey 2020);  near-IR photometry of TiO implied 
a slightly lower temperature of 3520$\pm$25K (Harper et al. 2020).  Another analysis (Kravchenko et al. 2021)  suggests that temperature diagnostics 
from the near-IR  are to be preferred.  Acceptance of  that  conclusion suggests that  the submm and near-IR measures 
indicate a lower temperature than usual was present in the stellar photosphere.  Additionally, after passage of the dense material, the  electron density
in the southern hemisphere of the chromosphere when measured in  2020 February,  had returned to the lower  values found in 2019 January  (Dupree et al. 2020b).

Millimeter and centimeter  observations with the VLA (Matthews \& Dupree 2022)   in 2019 August  indicated substantially lower temperatures than usual: 2267K and 2583K.  
It has been known for
a long time that the outer atmosphere of Betelgeuse is irregular and inhomogeneous, containing both  hot and cold elements. In 1996, the 7mm radiation indicated 
a mean radius of 43.5 mas (Lim et al. 1998)  which lies at 2 R$_{star}$, clearly dispersed within warmer chromospheric plasma.   Other models (Harper et al. 2001) suggest
formation closer in at 1.2R$_{star}$.  This region could lie near
the temperature minimum above the photosphere, which semi-empirical models  (Lobel \& Dupree 2000) suggest 
is located at $\sim$1.1R$_{star}$, and `below' the chromosphere.  With this interpretation, the event had passed
through the submillimeter region in 2019 August leaving a cooler, less dense atmosphere.   If however, we adopt a model with  columns of warm and cool material,  and the
warm regions are partly evacuated by the ejection event as indicated from ultraviolet observations, such an event would most likely also cause an expansion
 and contribute to a decrease in temperature  of the remaining  atmospheric material.    The passage of a shock wave could lead to different effective temperatures at 
 different atmospheric layers (Matthews \& Dupree 2022).

{\it This pattern of variation and atmospheric response suggests that an ejection event originated in the photosphere,  traveled out through the extended 
atmosphere, eventually creating molecules and dust, as cooler regions of the extended atmosphere were reached.  A  disturbed local  atmosphere, cooler, and of
lower density remained.  This would be expected when material is suddenly removed from the photosphere allowing  the remaining plasma
to expand and cool.}

\medskip
\section{Solar Coronal Mass Ejections}
It is helpful to review the phenomena known as {\it coronal mass ejections} (CME) that occur in the Sun, typically associated with 
solar flares.  
Solar coronal mass ejections  result from an impulsive ejection  of material, generally from the solar corona  into the heliosphere  (Tian \etal\ 2021).   These 
plasma evacuations  result in  CME-induced coronal `dimmings' - a weakening of coronal emission  - which is attributed primarily to a decrease in the coronal 
density. The density decrease can last for hours following the CME (Veronig \etal\ 2019).   Cases exist where the 
chromosphere is affected by a dimming as well.  For instance the  chromospheric He I, 10830 \AA\ line can respond  in addition to  the material at  coronal 
temperatures (Jiang \etal\ 2007) .  Study of plasma diagnostics following the event reveals additional significant  effects.  Construction of a differential 
emission measure  using many ions indicates a drop in the coronal 
temperature of 5-25\% occurs  along with the decrease in density ranging between 50-75\% (Vanninathan et al. 2018).   This solar process can inform 
the events observed on Betelgeuse.  

\section{A Surface Mass Ejection (SME)}

By contrast with the Sun, Betelgeuse possesses lower gravity (by a factor of $\sim$ 10$^4$), a substantially larger radius (by a factor of $\sim$ 1000), 
and lacks the high temperatures found in the solar corona (Dupree et al. 2005; Kashyap et al. 2020).  Although weak variable magnetic fields 
have been measured in the star (Auri\`ere et al. 2010; 
Mathias et al. 2018), there is no record of flaring as found in the Sun. 
However, Betelgeuse is known to have a fundamental pulsation period likely  driven by the $\kappa$-mechanism (Joyce et al. 2020),
large scale bright  variable convective cells on its surface  (Haubois et al. 2009 ; Kervella et al. 2018; Montarg\`es et al. 2016) which create chromospheric bright regions 
(Gilliland \& Dupree 1996; Dupree \& Stefanik 2013),  and asymmetric variable  large scale chromospheric motions (Lobel \& Dupree 2001).
Radiative hydrodynamic simulations also suggest the presence of large vigorous convective plumes in the photosphere (Freytag  et al. 2002, Chiavassa et al. 2009, Chiavassa et al. 2010,  Goldberg et al. 2022).

The sequence of observations discussed above suggests that Betelgeuse experienced a surface mass ejection initiated by the confluence of an extended
span of the outwardly moving photosphere, and the presence of  shock motions likely due to a vigorous convection cell.  These combined to 
eject  photospheric material through the chromosphere and into the extended atmosphere. 
  
\begin{figure}
\begin{center}
\includegraphics[angle =90,scale=0.75]{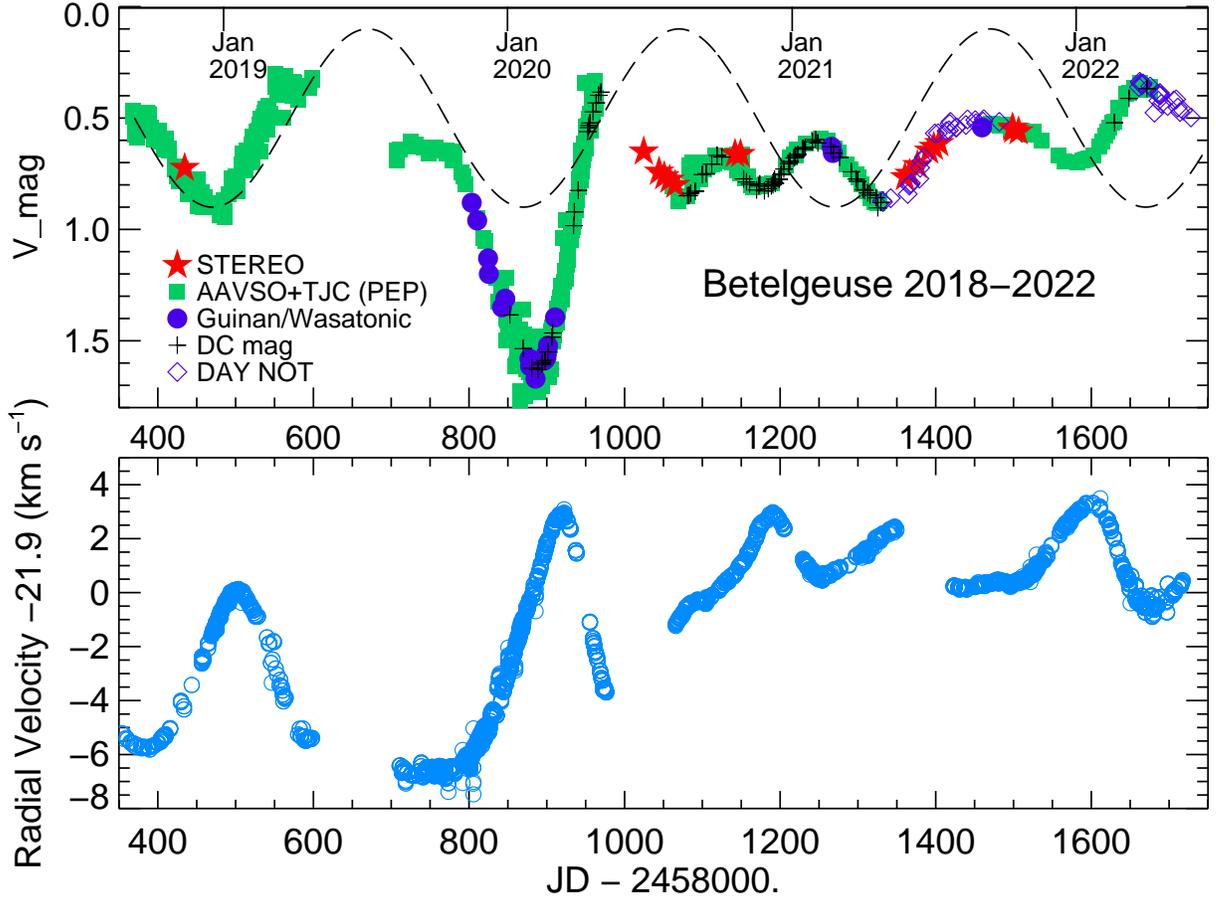}
\end{center}
\caption{V-Magnitude ({\it upper panel}) from the AAVSO (2022), STEREO, the Wasatonic Observatory, D. Corona, and daytime measures (DAY NOT). 
The broken line marks a  $\sim$400 day period.   This period is no longer present after the Great Dimming. {\it Lower panel:}  Radial velocity  of Betelgeuse from STELLA  during 2018-2022. }
\end{figure}

\newpage

\section{Betelgeuse 2021-2022: Atmospheric Response}

Following the Great Dimming, Betelgeuse continued to be monitored both photometrically  and spectroscopically (Figs. 4 and 5).  AAVSO members 
obtain frequent photoelectric photometer measures of the star, including some obtained during the daytime hours with multiple short exposures 
(Nickel \& Calderwood 2021).  Additionally,  the Solar Terrestrial Relations Observatory spacecraft (STEREO-A) located in the Earth's orbit but behind
the Earth was rolled to repoint to Betelgeuse during the summer and fall of 2020 and  2021.  The procedure is described elsewhere (Dupree et al. 2020a).  
An updated change in the sensitivity of the outer Heliospheric Imager HI-2, was used to extract the V-magnitude in 2021
relative to the measures in 2020 (Tappin et al. 2022).  These values agree well with the daylight measurements reported by the AAVSO.  

\begin{figure}
\begin{center}
\includegraphics[angle =90, scale=0.75]{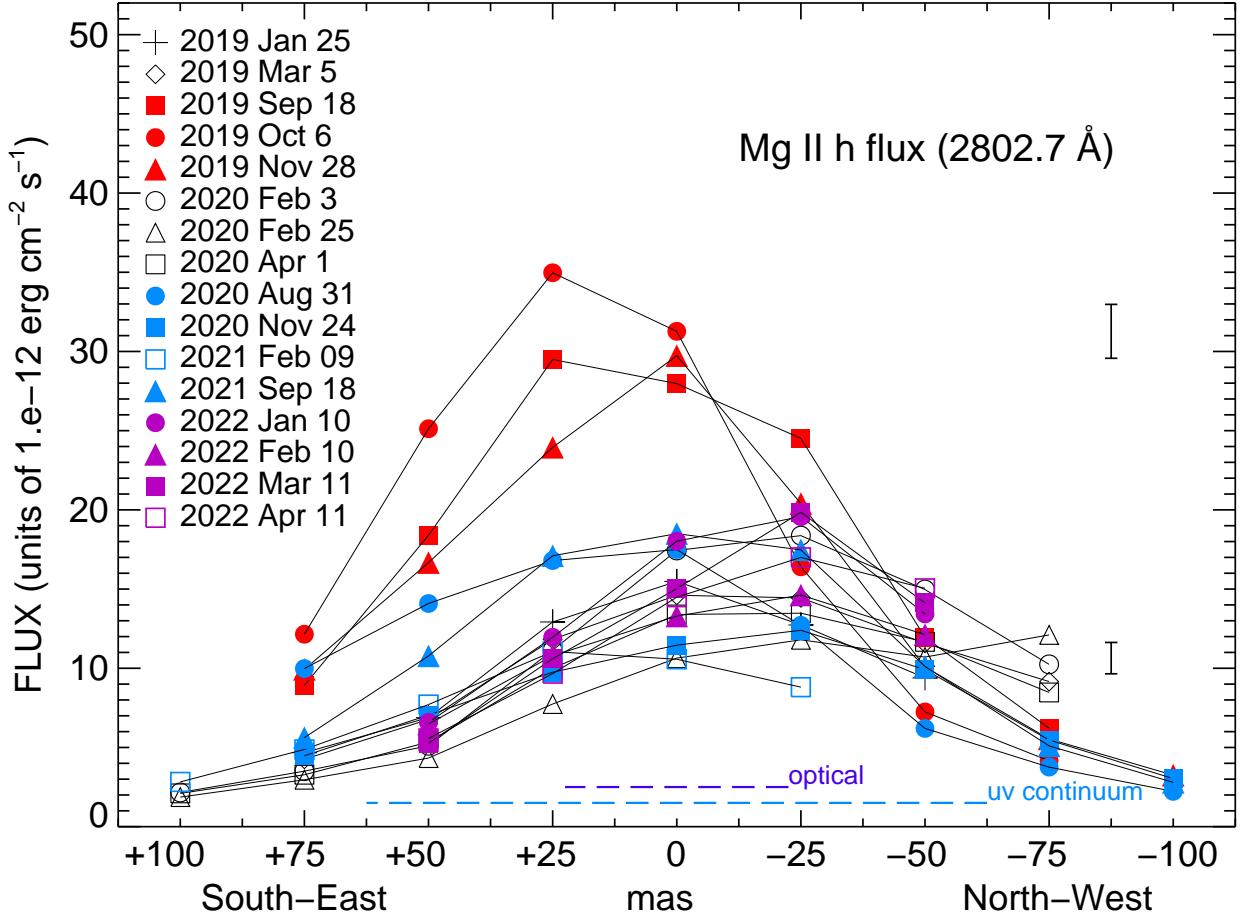}
\end{center}
\caption{Mg II-h flux obtained with STIS using a 25$\times$100 mas aperture, and various offset positions.  The aperture angle
varies among the observations but is generally in the quadrant marked on the x-axis.   The high chromospheric flux associated with the mass ejection event in
Sept-Nov 2019 has not been observed subsequently. In 2022, STIS pointings were made at 5 positions on the disk: $\pm$50 mas, $\pm$25 mas, in the center.
Optical and ultraviolet (FWHM) diameters are marked.  The error bars represent the uncertainly in the STIS flux measurement.}
\end{figure}

Inspection of the V-magnitude variation subsequent to the Great Dimming (Fig. 4) reveals dramatic changes from the 
well-documented $\sim$400 day pulsation (Kiss et al. 2006).  The periodicity
has shortened substantially.   After the Great Dimming, the  next minimum followed in $\sim$ 189 days, but  minima  in the following two years 
are separated by about  97 days, 166 days, and  233 days --all values clearly less than the dominant $\sim$400 day pulsation period.  Detailed study of 
periodicities in the V-magnitude spanning 88 years
 (Kiss et al. 2006) identified two dominant periods 388$\pm$30 days and 2050$\pm$460 days.  Photometric observations over 15  years 
indicated a fundamental period
 of 416$\pm$24 days (Joyce et al. 2020).  A short period of $\sim$ 200 days resulted from analysis of a 9.8 year 
study of magnitude variations (Percy et al. 1996).  The  surprising variable periodicity observed subsequent to the Great Dimming appears
to  have resulted from the   rearrangement  of the photosphere that followed the mass ejection and  which was subjected to the 
presence of the underlying fundamental 400-day pulsation.

Optical spectra continued to be  acquired with the STELLA echelle spectrograph (SES) on the robotic 1.2 m STELLA-II telescope located 
at the Iza\~na Observatory on Tenerife in the Canary Islands (Strassmeier et al. 2010).  These spectra span 390nm to 870nm with a 
spectral resolution of 55,000 (3-pixel sampling).  Standard reduction procedures were carried out with an IRAF-based pipeline 
(Weber et al. 2008; 2012).  Wavelength calibration was made using Th-Ar lamp spectra that are obtained consecutively.  After correction 
for the echelle blaze, radial velocities were derived using 62 of the 82 echelle orders.  The very short wavelength regions with weak 
signals were excluded along with orders containing very strong lines (e.g. Balmer series) and orders with heavy telluric contamination.   
Cross-correlations were made with a spectral template corresponding to 3500K, log g=0 and standard metallicity.  An order-by-order, 
(for 62 orders)  cross correlation utilized the  template spectrum  calculated with the Turbospectrum synthesis code (Alvarez \& Plez 1998) 
with MARCS models (Gustafsson et al. 2008)  and VALD3 line lists (Ryabchikova \etal\ 2015).  The cross-correlation functions are 
combined and the final velocity derived by fitting a gaussian curve to the peak.  The internal RV error is estimated with a 
Monte Carlo-like approach, using 1000 different subsets of the 62 orders and generally average to 10  m s$^{-1}$ (Granzer et al. 2021).

The photospheric radial velocity measured by STELLA (Granzer \etal\ 2021) also changed dramatically following the 
Great Dimming (Fig. 4). The amplitude is much smaller  ($\sim$ 5~km~s$^{-1}$) starting in 2020 September,  as contrasted 
with $\sim$10~km~s$^{-1}$ measured  previously, and the  outflow is minimal or absent.  The 
long-standing  pattern of maximum radial velocity inflow that follows  the optical minimum by $\sim$35 days for the past decade 
(Granzer et al. 2021)   is not present during 2020 August, but may have returned by 2020 December.  
Spatially resolved ultraviolet observations with the Hubble Space Telescope continue through 2020-2022 (Fig. 5) and the Mg II flux in 2022 
is considerably lower than 2019 September$-$November and appears unenhanced.

\section{Discussion}

The sequence of events described earlier suggests that Betelgeuse experienced a surface mass ejection. 
Consequences of the dimming event clearly affected the atmosphere of the star.  The optical variation following the dimming is dissimilar from period changes
found in giant stars such as Mira variables where period jitter of a few percent is common (Lombard \& Koen 1993), or the period may decrease systematically 
(Zijlstra et al. 2002), or mode-switching may occur (Bedding \etal. 1998).

Betelgeuse  undergoes substantial mass loss in a wind, and  clumps of dust in its environment suggest that material has been ejected  in  past events 
(Kervella et al. 2011; Humphreys \& Jones 2022).   Depending on assumptions, mass loss estimates causing the Great Dimming range from 3 to 128\% of the 
stellar wind mass loss rate (Montarg\`es et al. 2021).  These phenomena offer opportunities for mass loss in  both continuous and eruptive fashion.  
It will be of interest to search the circumstellar region to the South of the star to detect evidence of this mass loss event. At a nominal speed of 10 km s$^{-1}$, in 4 
years from the photospheric ejection in 2019  the material should be 38 to 48 mas distant from the stellar  surface  assuming pure radial motion on the sky and depending on 
the distance of Betelgeuse (Joyce et al 2020; 
Harper et al. 2017).  The apparent separation of the ejecta would be  larger than the  ALMA beam size at 18 mas at 0.88 mm (Kervella et al. 2018).

An analysis (Goldberg 1984) of  measurements of magnitude and radial velocity, led to the speculation that major disturbances in the atmosphere of Betelgeuse were 
likely to follow the minimum (maximum outflow)  in the ~6 year radial velocity variation.  This prediction of a disturbance in 1984 did not take place, and the
span of optical variability was  only 0.6 magnitudes   during that time (Percy et al. 1996).  The historic  optical minimum in 2020 did follow the minimum in the  long-term 
radial velocity variation (2229.8$\pm$6 days)  by approximately 300 days (Granzer et al. 2021).  This long period is expected to reach a minimum  again in May 2025, and perhaps in early 2026, 
another event may occur. 

\begin{acknowledgments}
This research is supported in part by HST/STScI grant HST-GO-15641 to the Smithsonian Astrophysical Observatory.
We would like to thank the STEREO/SECCHI science team at the Naval Research Laboratory and other locations, particularly Lynn Hutting, for setting up the special 
observing sequences, and the STEREO Mission Operations team at the Johns Hopkins University Applied Physics Laboratory for the special roll maneuvers needed 
to observe Betelgeuse while the star was unobservable from Earth due to solar interference.
\end{acknowledgments}

\facilities{AAVSO, FLWO(TRES), HST(STIS), STELLA(SES),  STEREO(HI)}

\end{document}